\documentclass[%
 reprint,
superscriptaddress,
 amsmath,amssymb,
 aps,
 pre,
]{revtex4-1}

\usepackage[brazil]{babel}
\usepackage[utf8]{inputenc}
\usepackage{graphicx}
\usepackage{dcolumn}
\usepackage{bm}
\usepackage{xcolor}
\usepackage[bf]{subfigure}
\usepackage{epigraph} 

\begin{document}

\title{Syntonets: Toward A Harmony-Inspired General Model of Complex Networks}

\author{Luciano da Fontoura Costa}
\email{Corresponding author: ldfcosta@gmail.com}
\author{Henrique Ferraz de Arruda}
\affiliation{ S\~ao Carlos Institute of Physics,
University of S\~ao Paulo, PO Box 369,
13560-970, S\~ao Carlos, SP, Brazil
}

\begin{abstract}
We report an approach to obtaining 
complex networks with diverse topology, here called \emph{syntonets}, taking into account the consonances and dissonances between notes as defined by scale temperaments. Though the fundamental frequency is usually considered, in real-world sounds several additional frequencies (partials) accompany the respective fundamental, influencing both timber and consonance between simultaneous notes. We use a method based on Helmholtz's consonance approach to quantify the consonances and dissonances between each of the pairs of notes in a given temperament. We adopt two distinct partials structures: (i) harmonic; and (ii) shifted, obtained by taking the harmonic components to a given power $\beta$, which is henceforth called the anharmonicity index. The latter type of sounds is more realistic in the sense that they reflect non-linearities implied by real-world instruments. When these consonances/dissonances are estimated along several octaves, respective syntonets can be obtained, in which nodes and weighted edge represent notes, and consonance/dissonance, respectively.   The obtained results are organized into two main groups, those related to network science and musical theory. Regarding the former group, we have that the syntonets can provide, for varying values of $\beta$, a wide range of topologies spanning the space comprised between traditional models.  Indeed, it is suggested here that syntony may provide a kind of universal complex network model. The musical interpretations of the results include the confirmation of the more regular consonance pattern of the equal temperament, obtained at the expense of a wider range of consonances such as that in the meantone temperament. We also have that scales derived for shifted partials tend to have a wider range of consonances/dissonances, depending on the temperament and anharmonicity strength.
\end{abstract}

\maketitle

\setlength{\epigraphwidth}{.49\textwidth}
\epigraph{``Non sai tu che la nostra anima è composta di armonia...''}{Leonardo da Vinci}

\section{Introduction}
Complexity corresponds to one of the most important subjects in science, which is often used to characterize not only real-world data but also several approaches to respective analysis and modeling. Not surprisingly, the particularly influential recent research areas of complex systems~\cite{costa2019quantifying} and network science~\cite{da2018complex,costa2011analyzing} are directly related to complexity. 

A great deal of the developments in network science is related to the systematic study of diverse connectivity patterns, as manifested in the topology of the respective networks. One of the most straightforward possible connectivity patterns consists of rings and lattices, characterized by full degree regularity,  and uniformly random networks such as Erdős-Rényi (ER)~\cite{erdos1959random} can be understood as stochastic instances of those perfectly regular structures.  At the other extreme of complexity, we have networks characterized by heterogeneous connectivity patterns, such as scale free~\cite{barabasi2009scale} and modular~\cite{fortunato2012community} networks.

Complex network topologies are particularly important because they can substantially influence dynamic processes related to changes in states associated with each network node (e.g.,~neuronal activation~\cite{lacoste2014sensory}) or on the network topology itself (e.g.,~resilience to attacks~\cite{motter2002cascade}).  Network complexity is directly associated to heterogeneity of one or more of its topological properties, such as node degree or clustering coefficient, among many others~\cite{costa2007characterization}.  For example, scale-free networks are characterized by a wide distribution of node degree, paving the way to the appearance of hubs ~\cite{barabasi2009scale}.

Several real-world structures --- such as protein interactions, airports, the internet, etc. --- are intrinsically heterogeneous so that many complex network models have been developed in order to explain and reproduce such specific real-world structures. Alternatively, complex network models can also be motivated by more mathematical approaches, such as the Apollonian circle packings~\cite{graham2003apollonian}, convex polytopes~\cite{grunbaum1967convex}, and non-Hamiltonian maximal planar graphs~\cite{nishizeki19801}, among other models derived from number theory~\cite{gross2004handbook}.

Both real-world and theoretically inspired networks typically have \emph{parameters} that influence their specific properties. For instance, in the case of the ER model, the parameter corresponding to the connecting probability will respectively control the density of connections in the result networks. Frequently, small changes in parameter values tend to induce relatively small modifications in the obtained structures.

The present work focuses on a potentially new type of networks, namely those derived from the concept of \emph{consonance} and \emph{dissonance} between two distinct sounds or notes~\cite{da2019self} along a given scale and temperament. This type of graphs is henceforth called \emph{syntonets}, acknowledging the important role of syntony 
between frequencies for consonance. More specifically, we consider Helmholtz's interesting approach to sound consonance, involving the comparison of the frequencies of the partials involved in each of the two sounds, in order to estimate the consonance and dissonance between pairs of notes. 

Every real-world sound signal includes, in addition to its \emph{fundamental} frequency, a sequence of \emph{partial} frequencies, which are ultimately responsible why the same note can sound different when played at two distinct types of musical instruments.  Helmholtz postulated~\cite{plomp1965tonal,da2019self} that the consonance/dissonance between two sounds, as perceived by human beings, would be related to the relative position of the involved partials. Furthermore, with dissonance arising when two simultaneous partials from different nodes are not identical but still are too close for proper discrimination by our aural perception. So, by comparing the several involved partials, it is possible to derive overall indices quantifying the resulting consonance and dissonance between any pair of notes.

Theoretically, while the sound produced by an ideal string would exhibit partials following the harmonic sequence, this is hardly the case in real-world instruments, where nonlinearities imply shifts from the harmonic relative positions.  For instance, the notes produced by a piano can exhibit relatively intense partial shifts, especially at both the frequency extremities (bass and treble)~\cite{rigaud2013parametric}.

In the Western musical tradition, sequences of notes are typically arranged in terms of respective \emph{scales}, in particular, those corresponding to two of the Greek modes, namely the \emph{major} and \emph{minor} diatonic scales.  Even if we focus on the major scale (which we will adopt henceforth, for simplicity's sake), a large number of variations will still exist. Is corresponds to the several \emph{temperaments} that have been developed along centuries, as the means to allow \emph{modulations} and \emph{transpositions} employed to provide diversity and variation during the musical experience.

The organization of notes according to a particular scale
(mode and temperament) gives rise to an immediate important
question: which of these notes will sound consonant or
dissonant when played together?  This issue is of particular relevance because it provides much of the basis for musical composition, in the sense that the consonant/dissonant properties of pairs of notes can be considered for producing specific effects during music composition and performance.

\begin{figure*}
\centering
\subfigure[$\beta = 1$]{\includegraphics[width=0.3\textwidth]{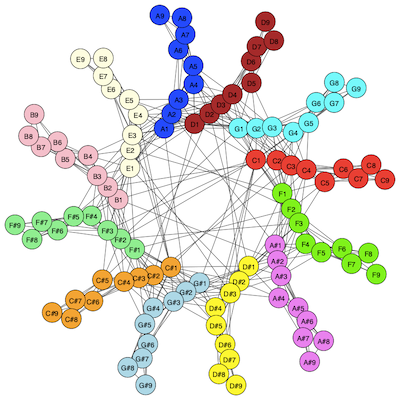}}
\subfigure[$\beta = 1.0056$]{\includegraphics[width=0.3\textwidth]{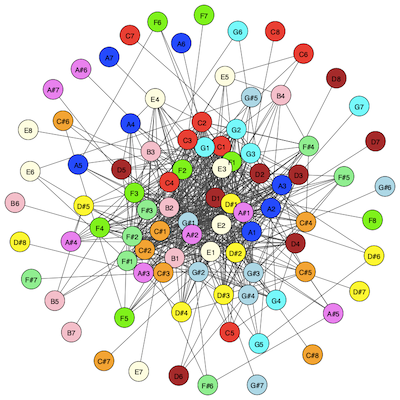}}
\subfigure[$\beta = 1.0058$]{\includegraphics[width=0.3\textwidth]{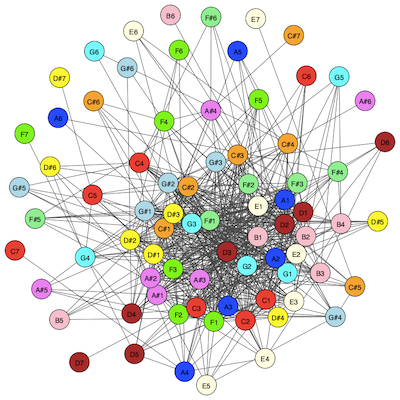}}
\caption{Example network of equal temperament, where items (a), (b), and (c) represent variations of $\beta$ parameter. Interestingly, a small difference of $\beta$ can lead to a substantial change in the network topology. }\label{fig:ex}
\end{figure*}

Helmholtz's approach to consonance provides an interesting means for quantifying, in an objective and automated manner, the pairwise consonance/dissonance of each of the pairs of nodes on a given scale~\cite{plomp1965tonal,da2019self}.  Now, if we represent each note as a node, and interconnect them according to the consonance/dissonance, syntonets can be obtained.  Figure~\ref{fig:ex}(a) illustrates one such network, corresponding to 9 octaves of the $C-$major scale under equal temperament, obtained by using the Helmholtz-inspired methodology considered in the current work while assuming the partials as being perfectly matched to the harmonic sequence. Figure~\ref{fig:ex}(b) also illustrates the same scale in (a), but now in the presence of partials shift, which is controlled by the anharmonicity parameter $\beta$.

This simple preliminary example illustrates interesting features exhibited by syntonets.  First, a remarkably regular pattern of consonance has been obtained, as shown in Figure~\ref{fig:ex}(a), with the notes having the same name defining more strongly interconnected groups or \emph{clusters}. Moreover, as it will be further discussed, many of the obtained consonances/dissonances tend to agree with the traditional harmonic view of interval consonances.  Striking results were also obtained regarding the effect of the partials shift, illustrated in Figure~\ref{fig:ex}(b).  More specifically, we have that such shifts can strongly influence the consonance/dissonance connecting patterns, in this specific case implying in making the network much less regular.  

Figure~\ref{fig:ex}(c) contains a third syntonet, obtained in the same way as the network in (b) but considering an almost identical anharmonicity index.  More specifically, the network in (b) was obtained  with $\beta = 1.0056$, while the structure in (c) used $\beta = 1.0058$.  So, interestingly the syntonets not only can be complex by themselves, but their properties can change drastically even for minute changes in the controlling parameters.

In the light of the remarkable properties of syntonets, as illustrated by only a few examples so far, it becomes interesting to investigate further the properties of this potentially new type of networks, stemming from musical motivation, which constitutes the main objective of the present work. More specifically, two main aspects will be approached: (a) to define and characterize syntonets from the perspective of network science research; and (b) to discuss musical interpretations and implications of the obtained results.

As it will be shown, syntonets are remarkably interesting because of their intrinsic potential to generate a wide diversity of topologies, especially in the presence of partials shifting controlled by the anharmonicity parameter $\beta$.  More specifically, the syntonets produced for varying intensities of this parameter cover almost completely the region comprised between more traditional models such as Barabási-Albert, Erdos-Rényi, Watts-Strogatz, stochastic block model, and spatial networks.  This interesting feature of syntonets is related to the number of theoretical aspects related to the distribution of partials along the frequency domain. We postulate that syntonets, especially considering a wider range of temperaments, could be understood as an \emph{overall generative model} capable of producing a substantially extensive range of topologies possibly encompassing those produced by more traditional models.

From the musical point of view, we will verify that the obtained results tend to agree substantially with the expected properties of the adopted temperaments. In addition, we will also observe that anharmonicity can significantly change the consonance/dissonance patterns induced by the considered temperaments.

This work starts by presenting the adopted methodology presenting the considered music concepts and types of temperaments, explaining the concept of partial structure, describing the methodology used to estimate consonance and dissonance. We explain how syntonets can be obtained, briefly describing the main measurements employed in order to characterize the topology of the obtained networks. Furthermore, outlining the important statistical method known as principal component analysis (PCA)~\cite{jolliffe2002principal}, which was used to obtain visualizations of the properties of the considered networks.  In the following section, results are presented and discussed regarding two main types of syntonets, namely respective to non-shifted and shifted syntonets. Conclusions and suggestions for future work to complete this article.

\section{Methodology}
In this section, we describe the employed concepts and methods, from the types of scales temperaments to PCA, also explaining the model that converts musical notes into networks and the adopted measurements that have been employed in the analysis of the respective topologies. 

\subsection{Some Basic Musical Concepts}

Western music has been organized around the concept of scales, modes, temperaments, and harmony.  In this section, we provide a concise introduction to the main musical concepts required for a better understanding of the reported methodology and discussion of the obtained results.  Thought the following concepts are those typically adopted in Western music, but also apply to many other musical traditions.

First, we have that each note has a characteristic pitch defined by the \emph{fundamental frequency of the sound}, namely the lowest frequency in its Fourier representation. For instance, the central A note typically is understood to be associated with the fundamental frequency of $440Hz$. Every real-world sound incorporates not only the respective fundamental component but also \emph{partials}, and it is this distribution of partials and respective intensities that characterizes the \emph{timbre} of the sounds.  In ideal instruments (linear), such as 
a string, the frequency distribution of these partials follow the harmonic series, i.e.~$f_1, 2f_1, 3f_1, \ldots$, where $f_1$ is the fundamental frequency.  However, in real-world sounds, the position of these partials are shifted as a consequence of non-linearities in sound production.  For example, a real string has a non-infinitesimal diameter, which implies non-linearities and partials shifting, especially at the lower frequencies, when the strings need to be much wider in order to keep the size of musical instruments manageable.  These frequency shifts are commonly related to the concept of \emph{anharmonicity}.

Sounds are typically organized into \emph{scales}, which have a characteristic mode, such as minor and major.  For instance, the C major scale is $C, D, E, F, G, A, B$.  Given 
any two notes on a specific scale, the \emph{interval} between them corresponds to the absolute difference, or ratios, of respective frequencies.  For instance, in the C major scale, the major fifth corresponds to the interval $[C, G]$.  The major fifth and the major third (i.e.~$[C,E]$) are critically important in defining the respective scale.  Indeed, the minor respective mode will mainly differ from the major counterpart with respect to the third, which is respectively minor and major.  Surprisingly, this small difference accounts for a great difference in the overall scale perception.   It is possible to \emph{transpose} any piece of music from a scale to another, simply by adding the frequency difference between the respective basic notes.  In order to account for diversity during the musical experience, it is also common to change the scale and mode along the same piece, which is called \emph{modulation} or \emph{progression}.

The specific assignment of given frequencies to the respective notes in a scale is provided by the chosen \emph{temperament}, to be further discussed in the subsequent section.  Of special notice is the \emph{equal temperament} as it paves the way to  
transpositions and modulations between any two scales and modes.

The simultaneous combination of two or more notes can induce varying perceptions on the
listener, and this depends on the interval between the involved notes.  For instance, two sounds with frequencies very similar tend to be perceived as being unpleasant.  Certain frequency relationships, such as in the octave, major fifth and major third, are typically perceived as being more \emph{consonant}.  Possible rules about how to combine sounds are the subject of the musical area known as\emph{harmony}.

\subsection{Types of Temperaments}
A scales can be defined as a sequence of notes, at subsequent frequency intervals, starting at a given node that gives the name to the scale. In this study, we focus on major scales starting at the $C1$, $f_1 = 32.7032Hz$. 

The actual frequencies of a given scale, let's say $C1$ major, are specified by the respective \emph{temperament}.  Therefore, the choice of a temperament defines the actual extent of \emph{intervals} between any pair of notes belonging to a given scale. In this study, we employ five different temperaments including \emph{Equal}, \emph{just}, \emph{Meantone}, \emph{Pythagorean}, and \emph{Werckmeister}. In the case of the Equal temperament, the frequency ratios are computed as follows
\begin{equation}
r_i = 2^{\frac{i-1}{12}},
\end{equation}
Table~\ref{tab:scales} illustrates the employed frequency ratios for each of the adopted temperaments.

\begin{table*}[]
\begin{tabular}{|l|l|}
\hline
Temperaments & Frequency Ratios  \\ \hline
Equal        & 1.0000, 1.0595, 1.1225, 1.1892, 1.2599, 1.3348, 1.4142, 1.4983, 1.5874, 1.6818, 1.7818, 1.8877  \\
Just         & 1.0000, 1.0417, 1.1250, 1.2000, 1.2500, 1.3333, 1.4062, 1.5000, 1.6000, 1.6667, 1.8000, 1.8750 \\
Meantone     & 1.0000, 1.0449, 1.1180, 1.1963, 1.2500, 1.3375, 1.3975, 1.4953, 1.5625, 1.6719, 1.7889, 1.8692 \\
Pythagorean  & 1.0000, 1.0535, 1.1250, 1.1852, 1.2656, 1.3333, 1.4238, 1.5000, 1.5802, 1.6875, 1.7778, 1.8984 \\
Werckmeister & 1.0000, 1.0535, 1.1174, 1.1852, 1.2528, 1.3333, 1.4047, 1.4949, 1.5802, 1.6704, 1.7778, 1.8792 \\ \hline
\end{tabular}

\caption{Frequency ratios characteristics of the five considered temperaments.  By considering a specific initial node defining the scale, the subsequent notes are obtained by multiplying the frequency of that note by the indicated values.  Observe that the values in this table are at lower resolution (fewer digits) that the values used in the reported experiments.}\label{tab:scales}
\end{table*}

The \emph{Pythagorean temperament} is one of the earliest approaches to defining the pitch of scales.  Its main characteristic is to have all fifth intervals perfectly aligned (and therefore consonant) at the expense of reducing the consonance between other intervals such as the third.  The \emph{just temperament}, which is also very old, tends to have an opposite effect as observed for the Pythagorean counterpart. The \emph{meantone temperament} applies mostly for keyboard instruments, dating back to early 1500.  Again, major thirds are defined to be more consonant, as well as several other intervals, so that a more balanced consonance is achieved along with the initial modes (e.g.,~C major, A minor, G major, among others).  However, transpositions or modulations to further away tones typically lead to considerable dissonances. The \emph{Werckmeister temperament} was introduced by Andreas Werckmeister in 1691, presenting three alternative systems, being oriented to organ tuning~\cite{werckmeisterWebsite,HARPSICHORD}.  This temperament preserves the perfect fifths while favoring transposition and modulations. The \emph{equal temperament} has been considered for a long time in western music, being more systematically formulated around 1600, and consolidated by composers such as J. S. Bach.  This system is especially important as it implies that the extent of intervals is preserved at any possible tonality.   In particular, the semitone ratio corresponds to $\sqrt[12]{2}$. It is achieved by increasing the 
dissonances of some intervals, such as the major third, but now the effects of transposing or modulating to any possible tonality are all perfectly equivalent, favoring these transitions.

\subsection{Harmonic Frequencies}
Taking the harmonic sequences in a string, defined as $l = \frac{1}{1}, \frac{1}{2}, \frac{1}{3}, \frac{1}{4} \dots$, the respective frequencies are given by $h_i = 1, 2, 3, 4, \dots$, where $f_1 = 1$ if the \emph{fundamental frequency}, and the remaining terms are known as \emph{harmonic partials}. The complete sequence that defines a sound is given by 
\begin{equation}
    f_1 h_1,  f_1 h_2,  f_1 h_3, \dots,  f_1 h_N,
\end{equation}
where $f_i h_i$ represents the harmonic partials. More specifically, the frequencies of the harmonics can be calculated as 
\begin{equation}
    f_n = f_1 h_n.
\end{equation}

We also consider non-harmonic sounds, more specifically those defined as above, but subjected to an anharmonic transformation implemented by the following power operation:
$h_n$ as
\begin{equation}
    h_n = n^{\beta},
\end{equation}
where $\beta$ induces a partial shift, implying a respective deviation from the respective harmonic partial. In the specific case of $\beta = 1$, we recover the perfectly harmonic partials. Strictly speaking, a harmonic component can be understood as a partial, but not vice-versa.

In order to find the highest frequency to be taken into account, we considered the maximum limit of audible frequency, which is 20,000Hz. So, the maximum $n$ was respectively found.

It is also necessary to specify the amplitudes of these partials, and here we adopt an exponential decay, as follows
\begin{equation}
    M_n = \exp{(-\alpha * h_n)},
\end{equation}
where $\alpha$ is a constant that controls the decay.

\subsection{Approaches to Consonance and Dissonance}
Helmholtz believed that if two tones had frequencies with minute differences, these tones could be considered as being consonant~\cite{plomp1965tonal,da2019self}. Similarly, Helmholtz considered dissonant frequencies being similar but less than in the consonant cases. Because there are divergent theories regarding consonance and dissonance, some researchers studied approaches that take into account human perception~\cite{plomp1965tonal,guthrie1928fusion}. Guthrie and Morrill~\cite{guthrie1928fusion} analyzed and compared the human perception of \emph{pleasant} and \emph{consonant} sounds. On the basis of previous studies, Plomp and Levelt~\cite{plomp1965tonal} also examined people's perception by using a consonance score. Interestingly, they found a strong relationship between their results and Helmholtz's theory. More specifically, they showed that specific bandwidths define the relationship between consonant and dissonant sounds.

Other related aspects have also studied, which include the \emph{local consonance} and the relationship between consonant scales and consonant timbres~\cite{sethares1993local}. Furthermore, this author described that the choice of timbres, for a given set of tones, can be understood as a problem of optimization. More recently, \emph{Berezovsky}~\cite{berezovsky2019structure} employed statistical mechanics-based techniques to explore patterns in music. In particular, he proposed a framework that included theoretical formalism, mean-field approximation, and numerical simulations.

\subsection{Harmonic Connectivity}
We compute the consonance and dissonance from a given pair of notes, $X$ and $Y$. In the case of consonance, we compare each of the $X$ frequencies with all $Y$ frequencies, and we consider as consonants the frequencies within the interval  $\Delta_{Min}$. More specifically, the comparisons between a given frequency $X_i$ and all $Y$ frequencies is given by
\begin{equation}
    c_i = \sum_{\substack{j \text{ so that}\\ 
    | X_i - Y_j| < \frac{\Delta_{Min}}{2}}}{A(X_i)A(Y_j)},
\end{equation}
where $A(\cdot)$ accounts for the amplitude of a given partial. Finally, the overall level of consonance is computed as
\begin{equation}
    C = \sum_{i=1}^N{c_i},
\end{equation}
where $N$ is the number of harmonics of $X$. We consider a pair of notes as dissonant when their pairs of partials are close, but not as close as to be consonant, as follows
\begin{equation}
    d_i = \sum_{\substack{j \text{ so that}\\
    | X_i - Y_j| \geq \frac{\Delta_{Min}}{2}\\
    \text{and } | X_i - Y_j| < \frac{\Delta_{Max}}{2}
    }}{A(X_i)A(Y_j)},
\end{equation}
where $\Delta_{Max}$ defines the dissonant maximum interval. The overall level of dissonance is computed as
\begin{equation}
    D = \sum_{i=1}^N{d_i}.
\end{equation}

\begin{figure}[!h]
\centering
\includegraphics[width=0.32\textwidth]{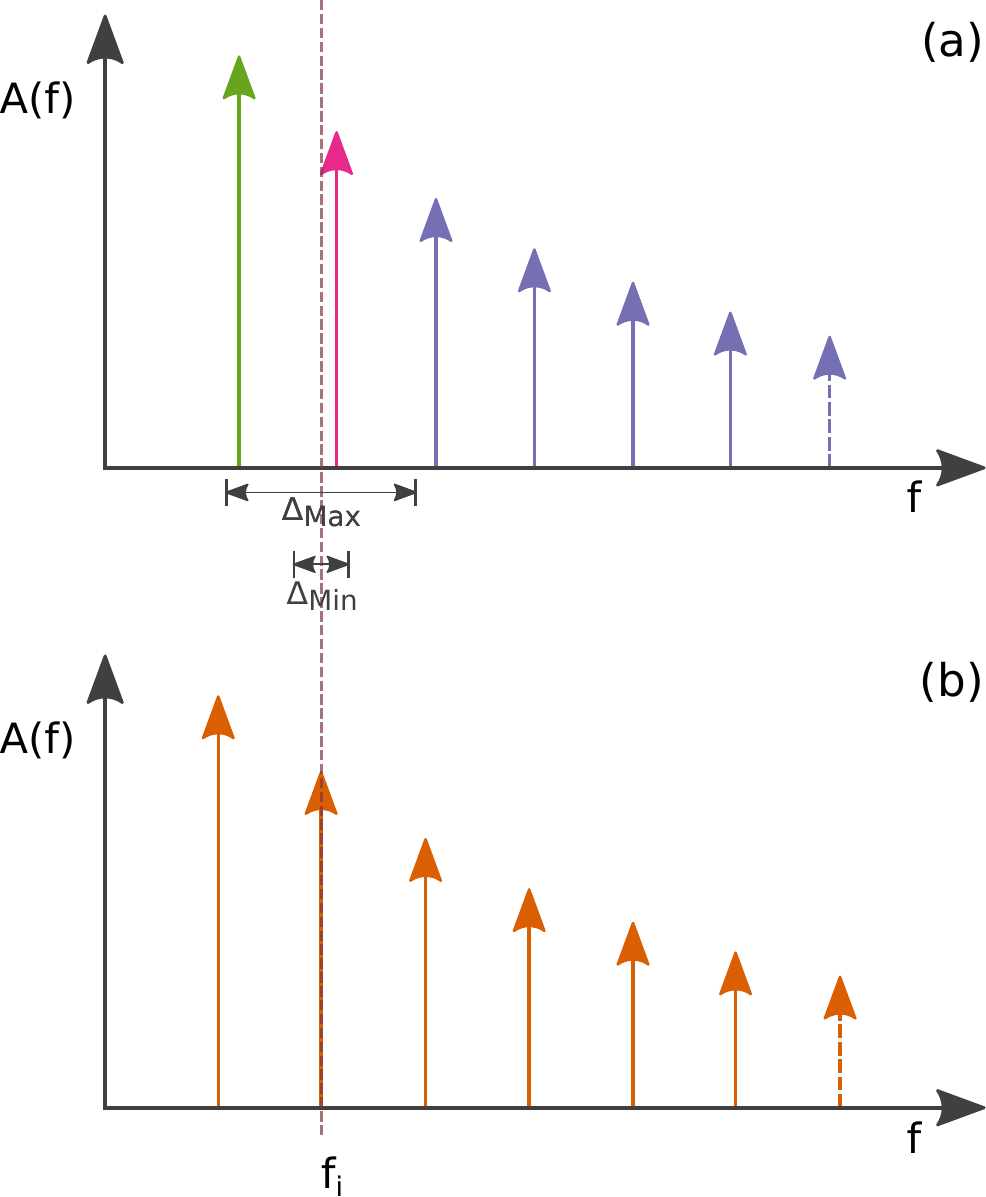}
\caption{Two notes (a) and (b) to have their consonance/dissonance quantified, with the latter being taken as reference.  For each partial $f_i$ belonging to the note (b), the partials of (a) comprised between the intervals $\Delta_{Max}$ and $\Delta_{Min}$ are identified. In this case, the pink and green frequencies of (a) are considered consonant and dissonant, respectively.}\label{fig:comparison}
\end{figure}

Figure~\ref{fig:comparison} illustrates an example of consonant and dissonant calculation. More specifically, a consonant frequency is within the interval defined by $\Delta_{Min}$. Furthermore, another frequency comparison is within $\Delta_{Max}$, but is not within  $\Delta_{Min}$, so it is considered dissonant.

By considering these estimated levels of consonance and dissonance, we created weighted networks. Each node represents a note, while the edges are weighted according to the consonance or dissonance levels between the respective pair of notes.  In order to allow the calculation of many of the existing complex network measurements~\cite{costa2007characterization}, we considered unweighted versions of the networks. More specifically, we removed the edges with the lowest weights aiming at having all the networks with the same number of edges. After this edge removal, we considered only the largest connected component. So, the number of nodes and edges can present some small deviations from the adopted reference values.

\subsection{Complex network measurements}
\label{sec:measurements}
Many distinct measurements have been proposed to grasp the characteristics of complex networks~\cite{costa2007characterization}. Here we considered the following measurements: Degree~\cite{costa2007characterization},  clustering coefficient~\cite{watts1998collective}, neighborhood~\cite{pastor2001dynamical} (undirected and directed, in and out, versions), accessibility~\cite{travenccolo2008accessibility} (for two and three concentric levels), generalized acessibility~\cite{de2014role},  backbone and merged symmetries~\cite{silva2016concentric} (for both versions we considered two, three, and four concentric levels), eigenvector centrality~\cite{bonacich1987power} (undirected version), betweenness centrality~\cite{freeman1977set} (directed version), edge betweenness centrality-directed~\cite{girvan2002community} (undirected version), and eccentricity~\cite{harry1969graph} (undirected version). For each of the computed measurement, we estimate the respective average and standard deviations.

In order to characterize the networks obtained in this work, we compare them with some reference, traditional models. Each of these models has specific topological features, as described in the following. The uniformly random model proposed by Erdős-Rényi (ER)~\cite{erdos1959random}, which is created from random connections, is defined by a single probability. As a consequence, most degrees result to be similar, approaching a regular network.  The Watts-Strogatz model (WS)~\cite{watts1998WS}, proposed as an approach to small-world networks, has a small average shortest path distance and high clustering (here we consider the initial network as a 2D toroidal lattice). Barabási–Albert approach (BA)~\cite{barabasi1999BA} consists of a network with scale-free degree distribution, favoring the existence of hubs. As a geographical model, we employ the Random geometric graph (GEO)~\cite{penrose2003random} (we set the initial positions of the nodes as a 2D lattice). The Stochastic Block Model (SBM)~\cite{holland1983stochastic} is a model incorporating communities. We configured this model to present two communities with the same size. In order to obtain more reliable statistics, we considered 100 network samples for each of the above models.

\subsection{Principal Component Analysis}
\begin{figure}
\centering
\includegraphics[width=0.45\textwidth]{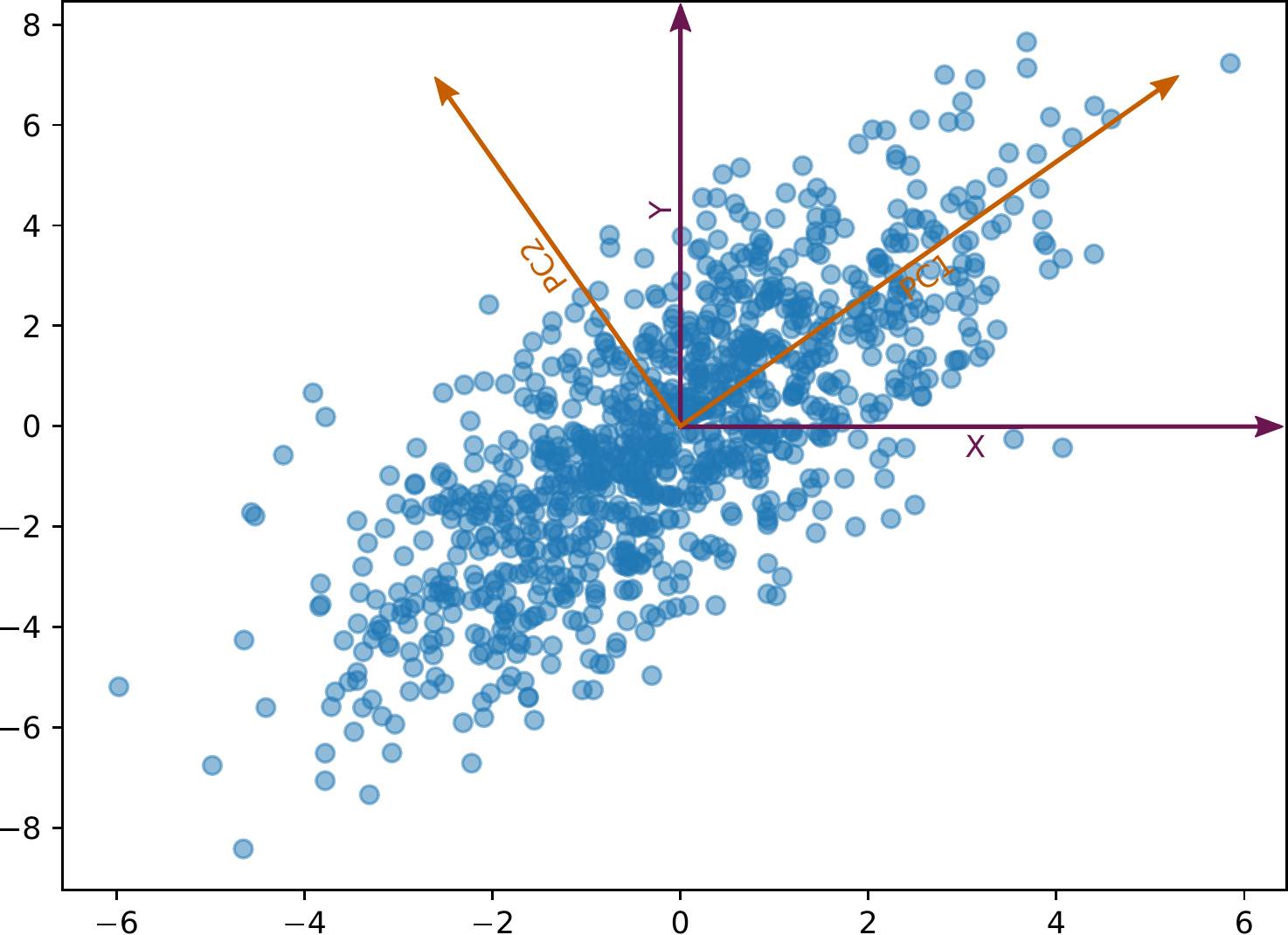}
\caption{Example of PCA for the data organized into two dimensions, in which the computed PCA axes reflect the dispersion of the data.}\label{fig:pca}
\end{figure}

Principal component analysis (PCA)~\cite{gewers2018principal,jolliffe2002principal} is an orthogonal transformation of the matrix of data $X$, in which each line consists of a vector of features of a given sample.  More specifically, the covariance matrix $K$ of the data $X$ is initially computed. From this matrix, the respective eigenvectors and eigenvalues are calculated. The projected axes are described by the eigenvectors that are ordered in decreasing order according to their respective eigenvalues, which describe the percentage of variance explanation provided by each subsequent axis.  In other words, this transformation represents a rotation, in which the transformed axes are ordered in decreasing order according to the data dispersion (see Figure~\ref{fig:pca}). One of the advantages of using PCA is that the projection completely decorrelates the involved random variables.  By allowing the dimensionality of the data to be reduced, it can provide valuable visualizations of the data dispersion in lower dimensions.

In order to better understand the characteristics of the proposed network representations, we compare their topologies with some well-known models. First, we measure many different network features from the proposed networks and network models. By considering only the models, we compute the PCA projections, which allow both types of networks to be visualized as 2D scatterplots. 

\section{Results and Discussion}
In this section, we present the obtained results and the respective discussion. The first part addresses the non-shifted partials ($\beta=1$), followed by the analysis of shifted partials. In both cases, the obtained syntonets are compared with some well-known traditional models. 

\subsection{Non-Shifted Partials}
First, we discuss the results obtained for the syntonets without shifting the respective partials so that all partials can be understood as being harmonics.  These networks are better appreciated when compared to more traditional complex network models, such as BA, ER, GEO, SBM, and WS.

\begin{figure*}[!htbp]
\centering
    \subfigure[Consonance]{\includegraphics[width=0.48\textwidth]{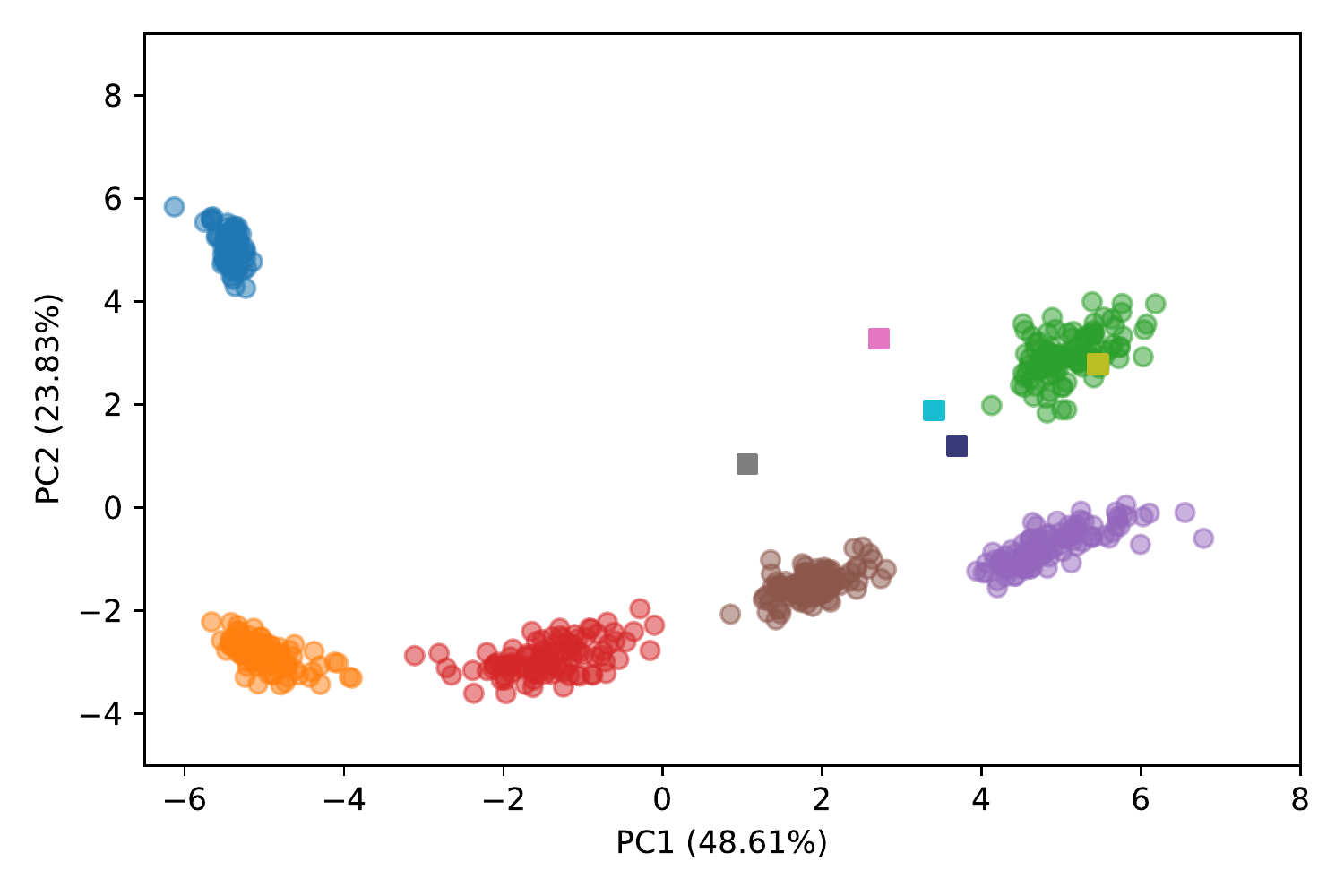}}
    \subfigure[Dissonance]{\includegraphics[width=0.48\textwidth]{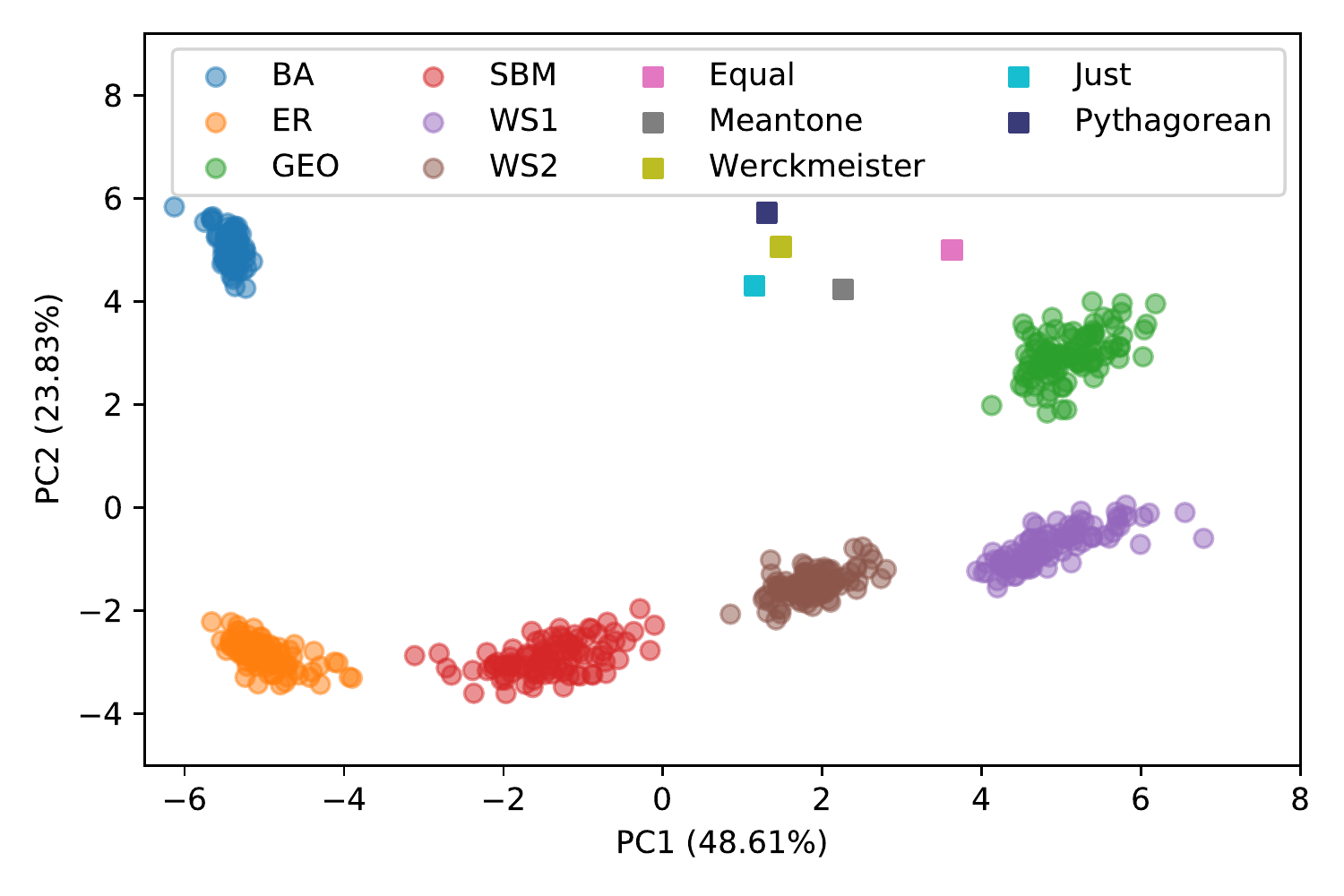}}
    
\caption{PCA projections respective to consonance and dissonance obtained for each of the considered temperaments, allowing a comparison of the non-shifted scale-based networks with some traditional network models.}
\label{fig:pcaNonShifted}
\end{figure*}

In order to perform such a comparison, we create 100 network samples for each of the models described in Section~\ref{sec:measurements}. The parameters of these models were chosen so that the average degrees resulted similarly to those of the considered syntonets. More specifically, $\langle k \rangle = 10.37$, which was chosen for the Equal temperament network through visual inspection. We calculated the network measurements (described in Section~\ref{sec:measurements}) for several instances of the considered traditional models and then obtained the respective PCA projection.  For all considered temperaments, both respective consonance and dissonance syntonets were projected using the same PCA matrix, yielding the overall map shown in Figure~\ref{fig:pcaNonShifted}.

As far as the consonance syntonets are concerned, we have that they resulted near the GEO, WS1, and WS2, with the Werckmeister network falling within the GEO cluster.

\begin{figure*}
\centering
\subfigure[Equal (consonance)]{\includegraphics[width=0.25\textwidth]{equalConsonance_1_0000.png}}
\subfigure[Meantone (consonance)]{\includegraphics[width=0.25\textwidth]{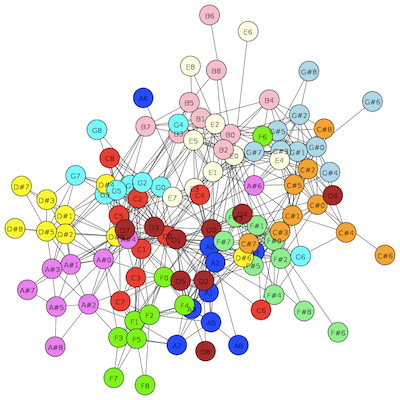}}
\subfigure[Werckmeister (consonance)]{\includegraphics[width=0.25\textwidth]{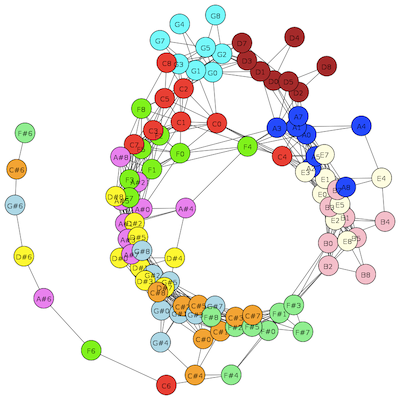}}
\subfigure[Just (consonance)]{\includegraphics[width=0.25\textwidth]{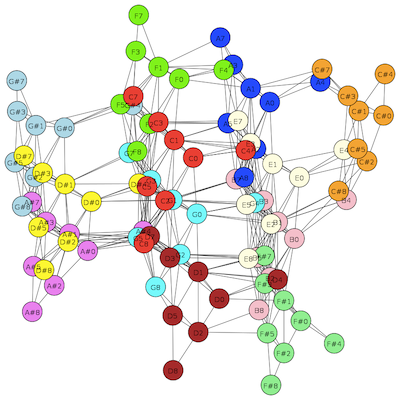}}
\subfigure[Pythagorean (consonance)]{\includegraphics[width=0.25\textwidth]{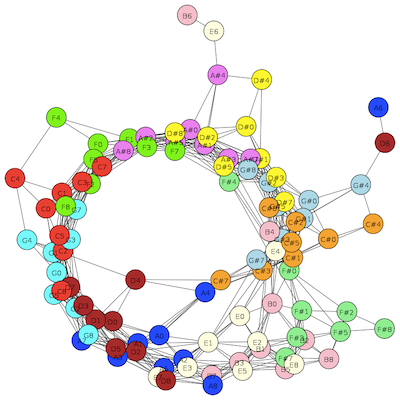}}

\vspace{1cm}

\subfigure[Equal (dissonance)]{\includegraphics[width=0.25\textwidth]{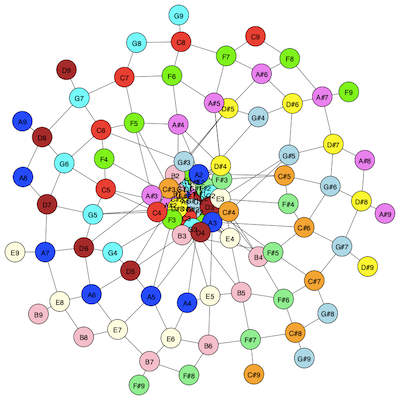}}
\subfigure[Meantone (dissonance)]{\includegraphics[width=0.25\textwidth]{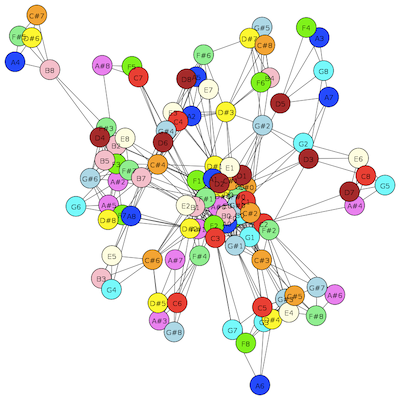}}
\subfigure[Werckmeister (dissonance)]{\includegraphics[width=0.25\textwidth]{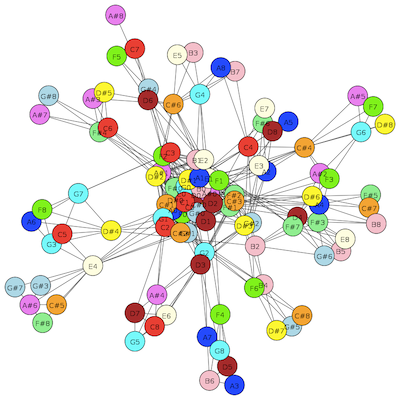}}
\subfigure[Just (dissonance)]{\includegraphics[width=0.25\textwidth]{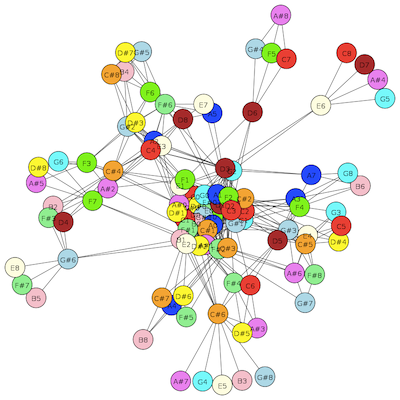}}
\subfigure[Pythagorean (dissonance)]{\includegraphics[width=0.25\textwidth]{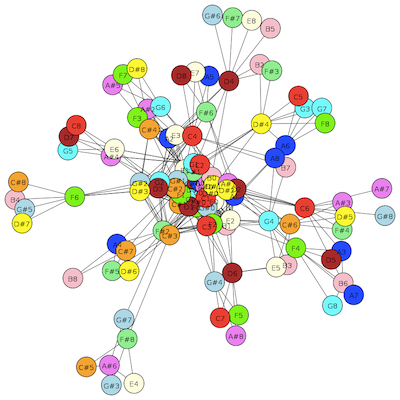}}
\caption{Visualization of consonance and dissonance networks we created from non-shifted partials, for all of the considered temperaments. \label{fig:networksVis}}
\end{figure*}

Figure~\ref{fig:networksVis} illustrates the syntonets obtained for each of the considered temperaments. The networks defined by the equal temperament (Figure~\ref{fig:networksVis}(a)) exhibits a well-organized, symmetric topology that is a direct reflex of the constant frequency intervals adopted in this type of scale.  However, the consonance is mostly concentrated between notes of the same name and intervals between $C$ and $G$ (specific fifth), with relatively minor consonances between most of the other intervals.  The meantone network (Figure~\ref{fig:networksVis}(b)) presents a more widespread and ample pattern of consonances, concerning several intervals, which is in agreement with its original musical motivation (aimed at achieving more consonances at the expense of transposition and modulation).  Both the Werckmeister and Pythagorean networks (Figure~\ref{fig:networksVis}(c)~and~(e)) present circular patters of consonances, but respective to different intervals.  The syntonet obtained for the just temperament exhibits a remarkably symmetric topological organization, suggesting a `crystalline' structure (Figure~\ref{fig:networksVis}(d)).  However, the consonance is again found between notes of the same type.

As shown in Figure~\ref{fig:networksVis}, the patterns of predominant dissonances varies substantially among the considered temperaments, but a more dissonant core can be observed in all cases.  As expected, the more symmetric, and also widespread, the structure can be observed in the equal temperament (Figure~\ref{fig:networksVis}(f)).  In the other 4 cases, the dissonances are less widespread among the notes outside the central core.

\subsection{Shifted Partials}
\begin{figure*}[!htbp]
\centering
    \subfigure[Equal]{\includegraphics[width=0.48\textwidth]{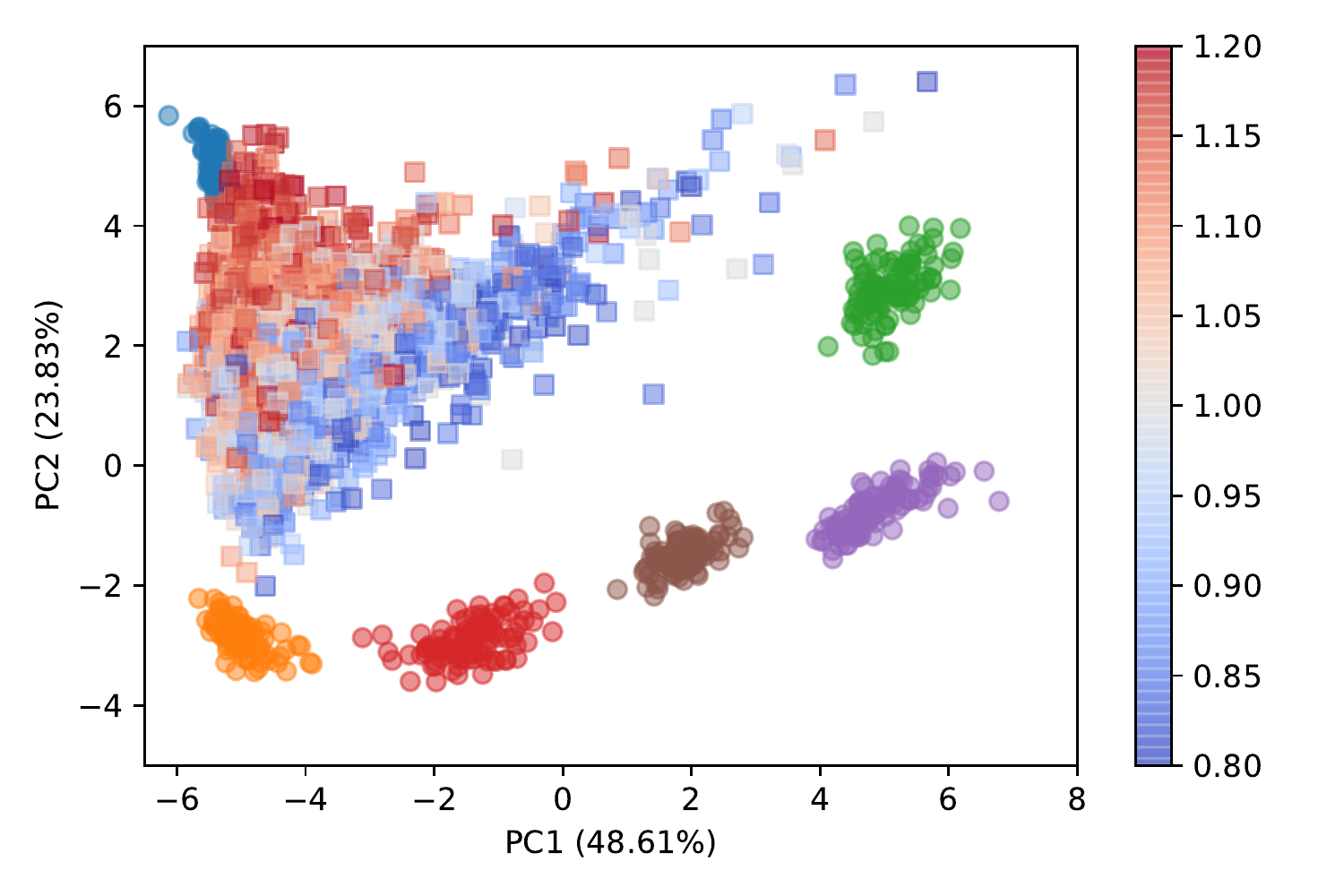}}
    \subfigure[Just]{\includegraphics[width=0.48\textwidth]{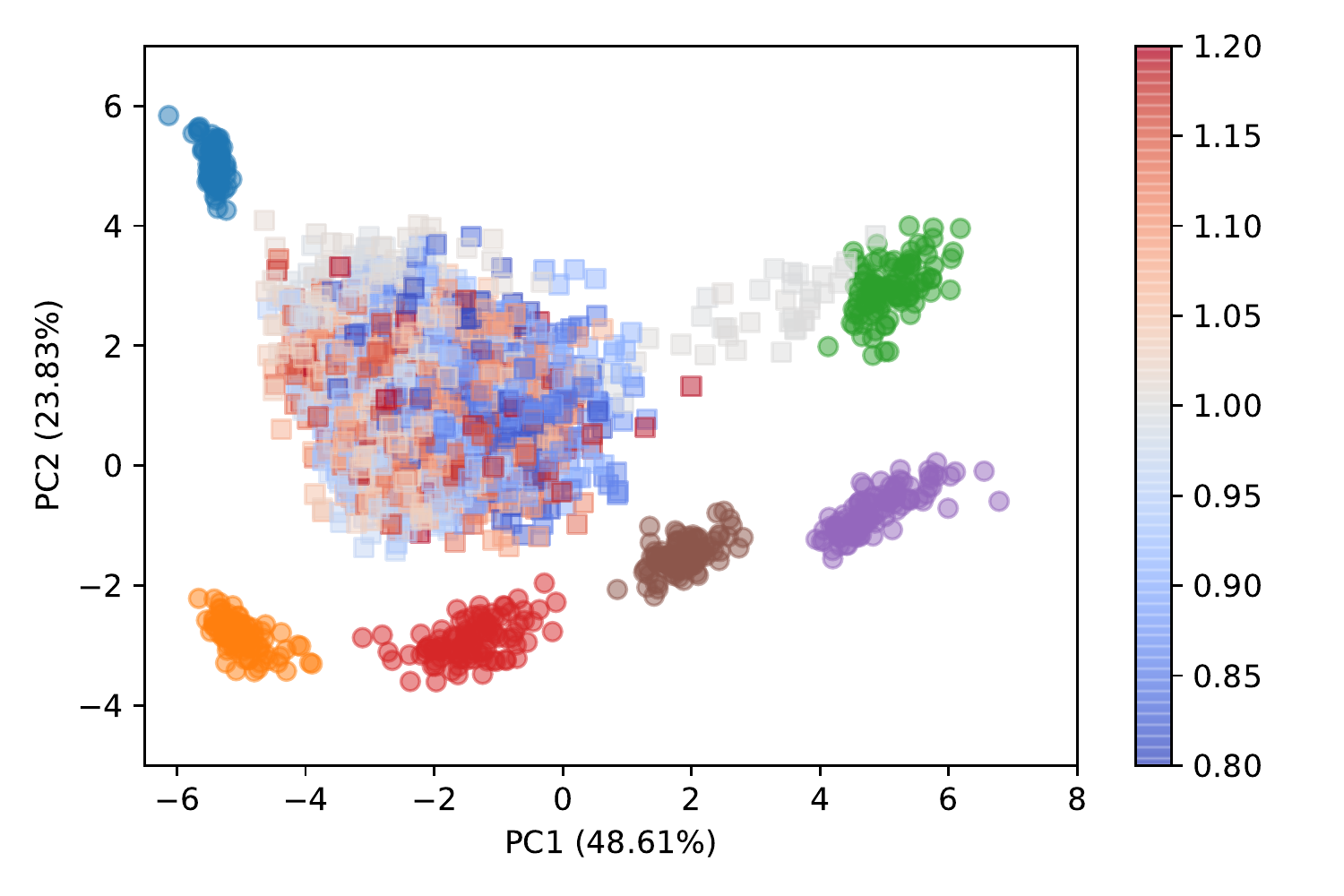}}
    \subfigure[Meantone]{\includegraphics[width=0.48\textwidth]{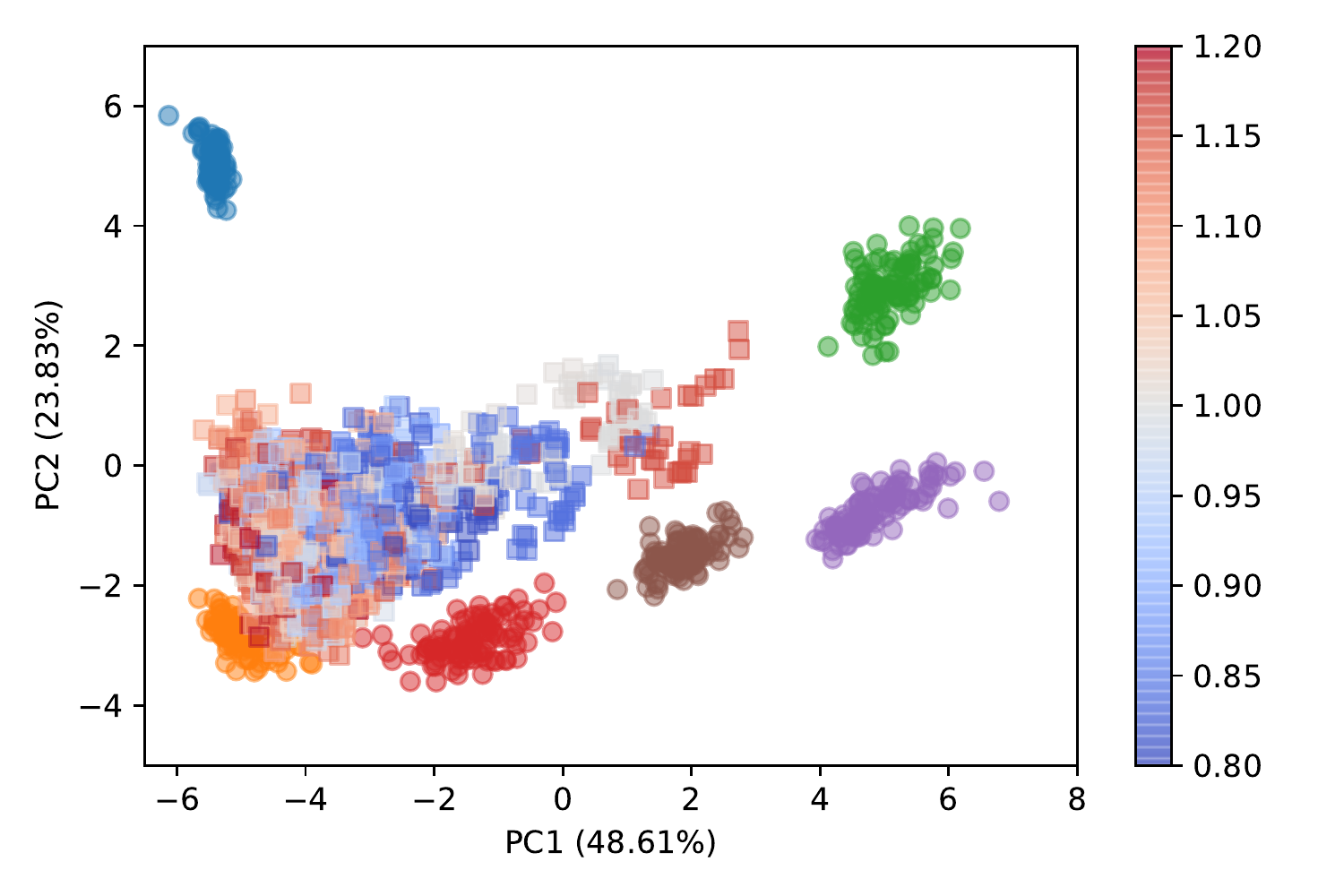}}
    \subfigure[Pythagorean]{\includegraphics[width=0.48\textwidth]{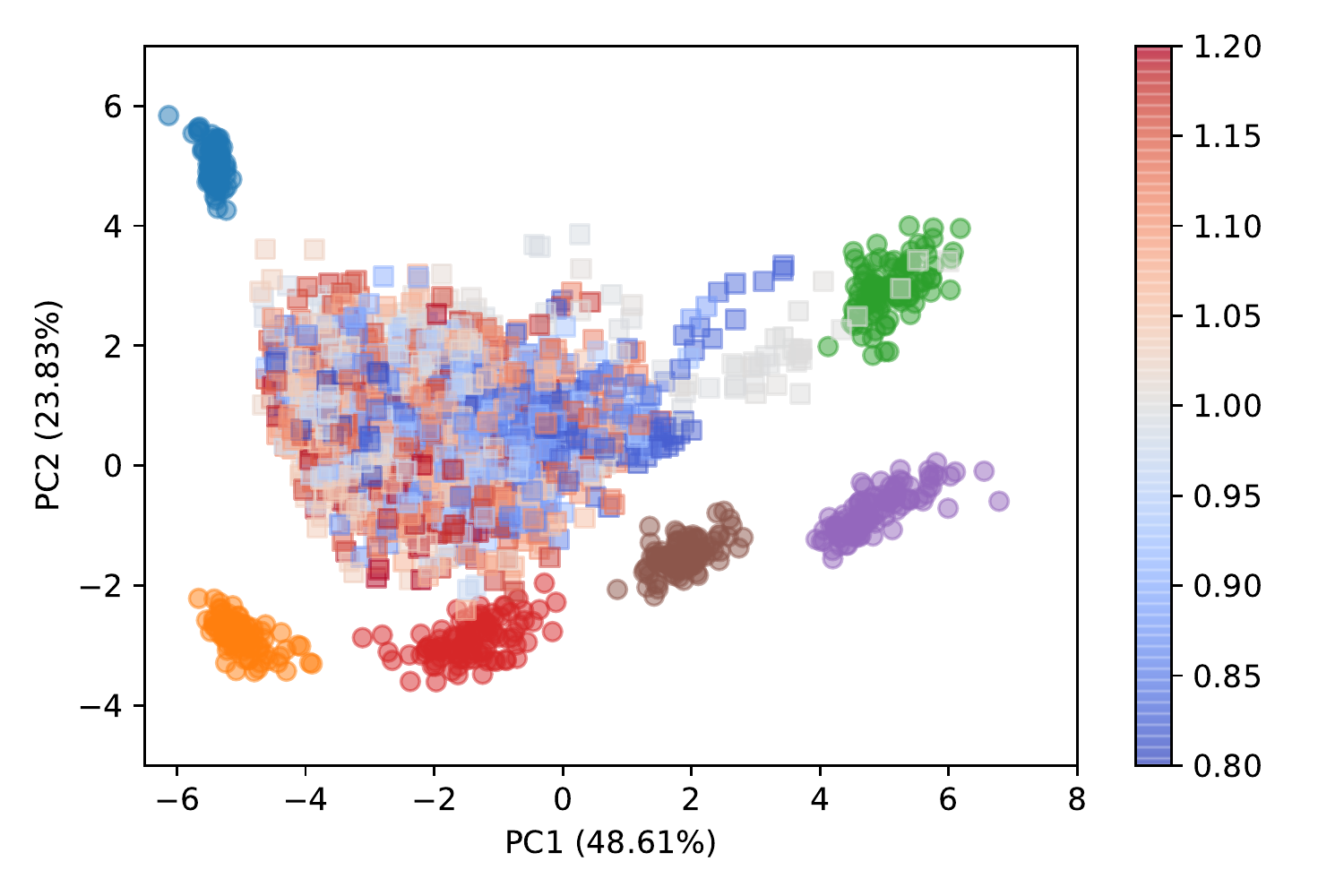}}
    \subfigure[Werckmeister]{\includegraphics[width=0.48\textwidth]{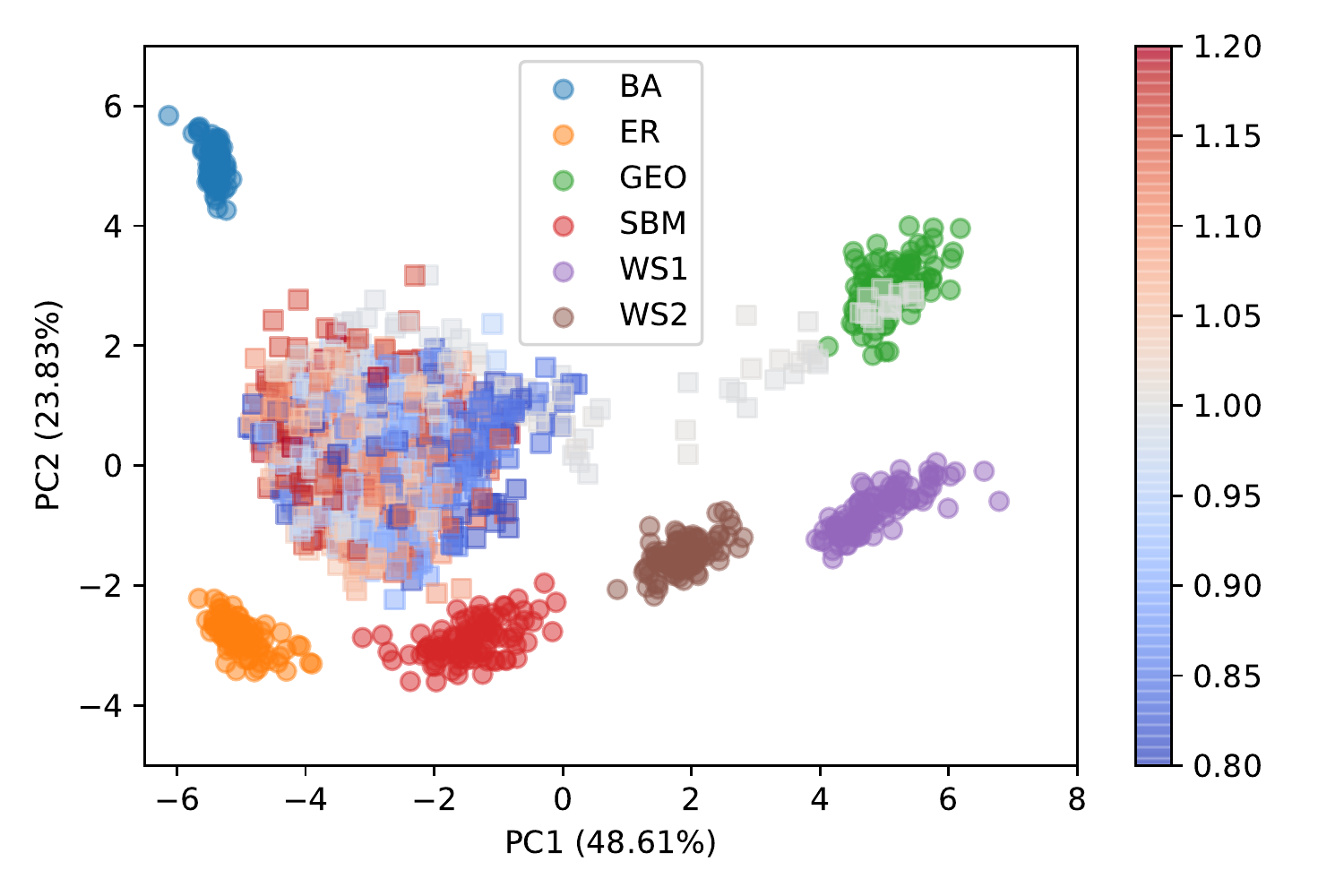}}
    
\caption{PCA projection of the consonance networks (with shifted partials) for all of the considered temperaments (squares). The circles represent the network models. \label{fig:pcaConsonanceShifted}}
\end{figure*}

Now we turn our attention to syntonets obtained in the presence of shifted partials, i.e., presenting an anharmonicity level determined by the parameter $\beta$.  Figure~\ref{fig:pcaConsonanceShifted} illustrates the PCA projections, respective to each temperament, obtained for the \emph{consonance} shifted-partial together with the more traditional complex networks considered in this work. The color bar indicates the values of $\beta$, which controls the shifting effect.

Remarkably, in all cases (i.e., temperaments), the syntonets tended to span the space comprised within the considered traditional complex network models.  

In the case of the equal temperament (Figure~\ref{fig:pcaConsonanceShifted}(a)), the obtained syntonets approach the BA model in the PCA space as $\beta$ increases from $0.80$ to $1.20$.  The effect of $\beta$ in defining the progression of these syntonets is evident in Figure~\ref{fig:pcaConsonanceShifted}(a).  This effect is substantially less prominent in the other temperaments, with the networks exhibiting substantial topological changes as $\beta$ is progressively increased.  The just, Pythagorean, and Werckmeister temperaments yielded the syntonets closest to the middle of the space defined by the more traditional complex network models.  The meantone temperament generated syntonets overlapping the ER model, therefore suggesting more uniform topological properties.

\begin{figure*}[!htbp]
\centering
    \subfigure[Equal]{\includegraphics[width=0.48\textwidth]{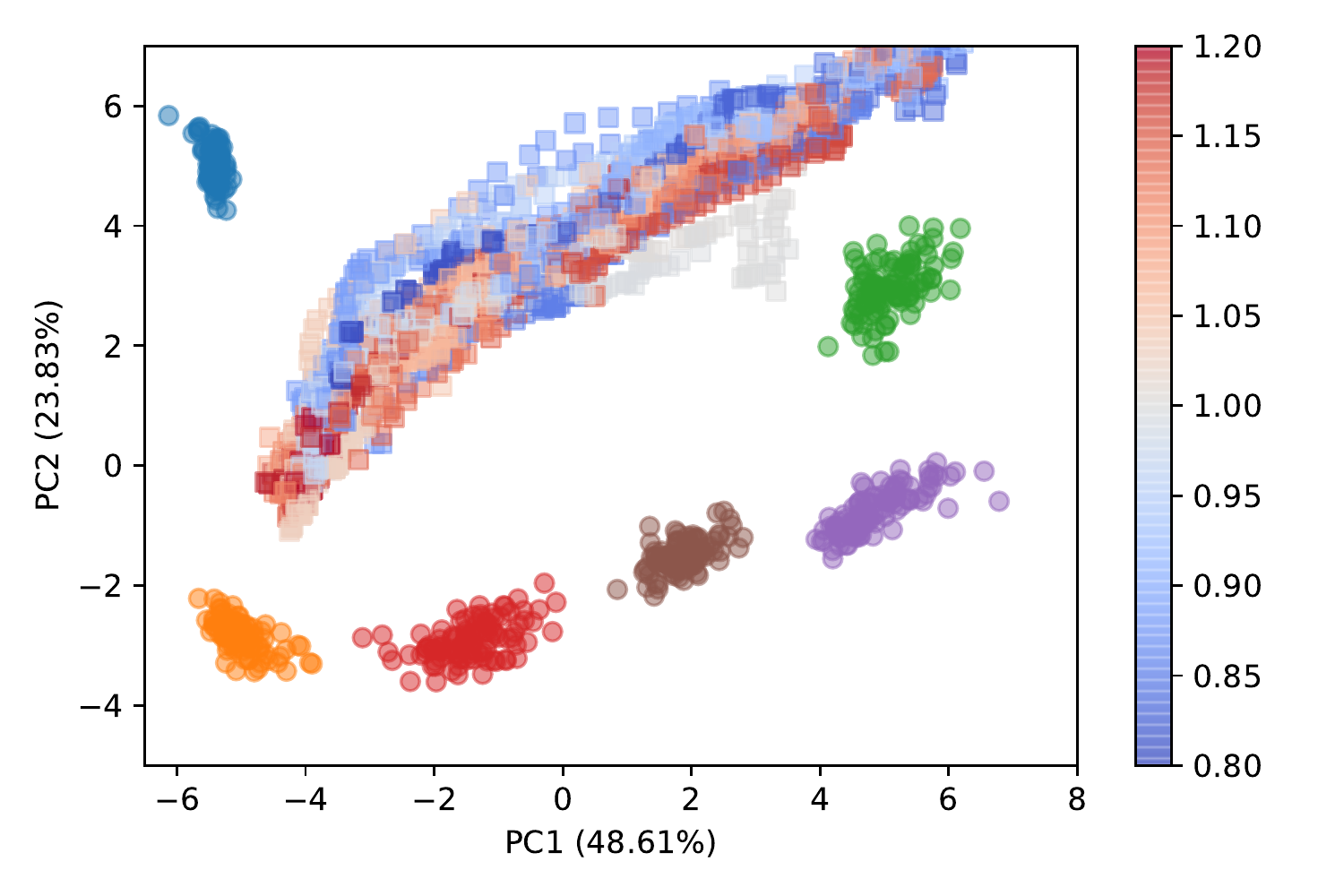}}
    \subfigure[Just]{\includegraphics[width=0.48\textwidth]{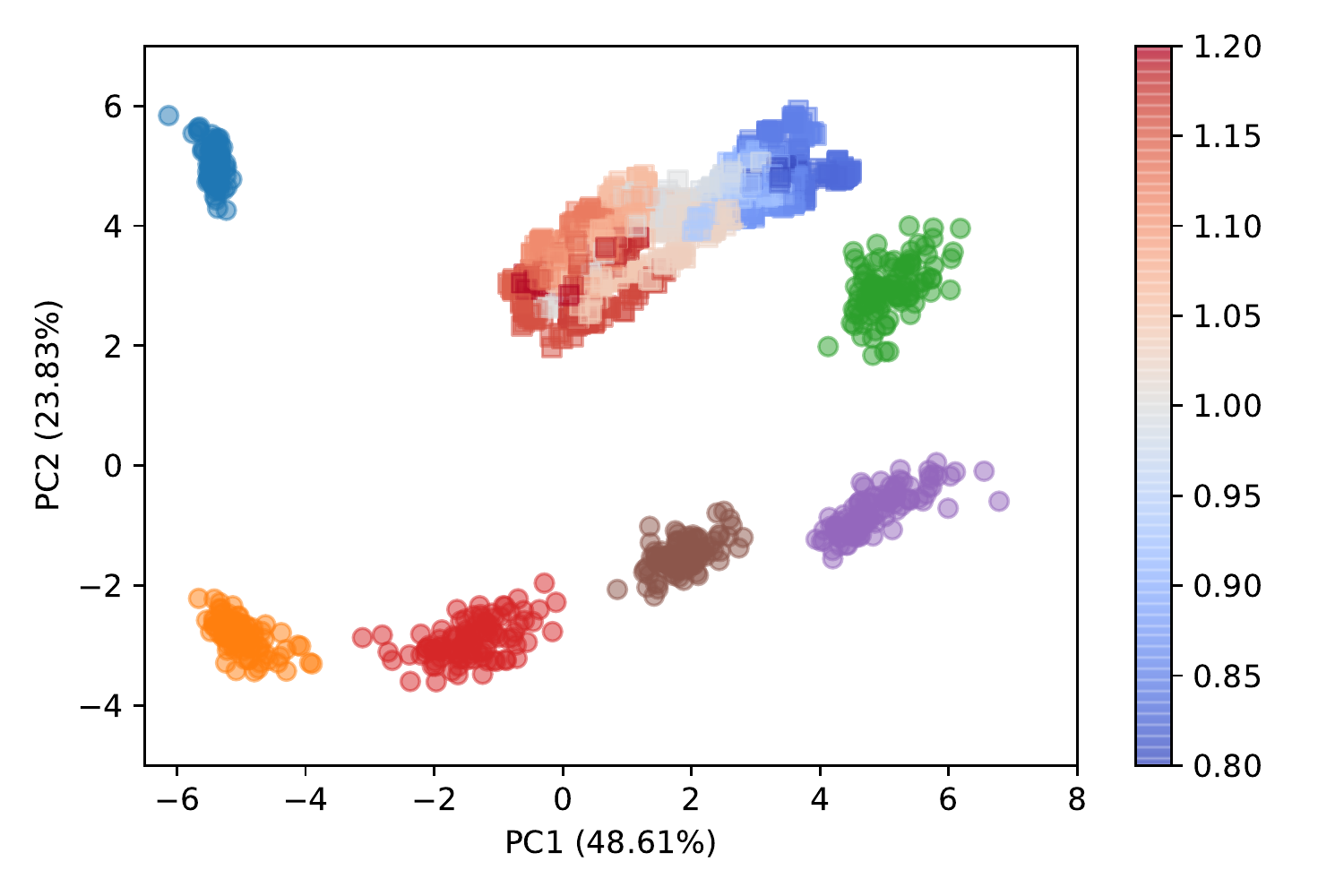}}
    \subfigure[Meantone]{\includegraphics[width=0.48\textwidth]{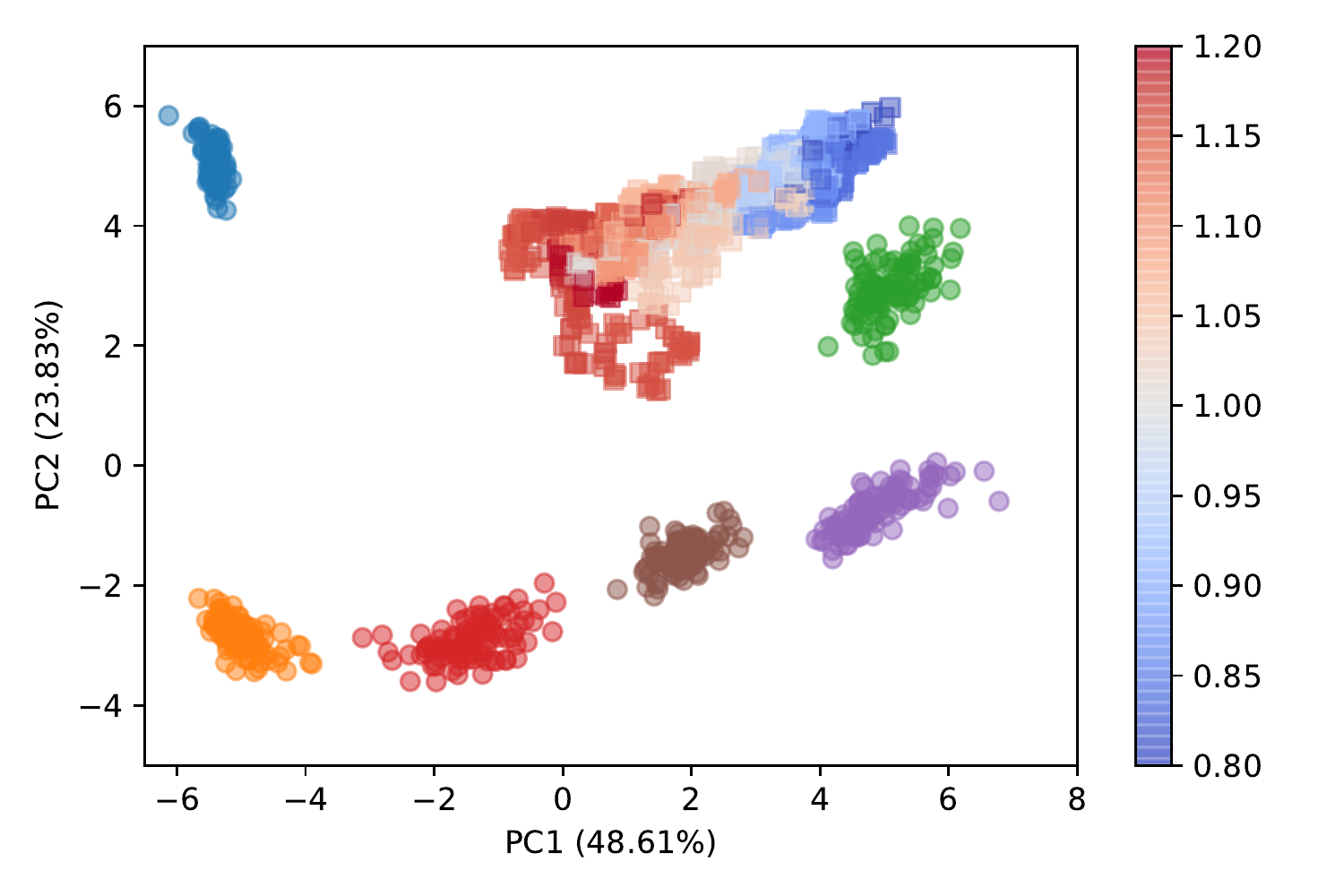}}
    \subfigure[Pythagorean]{\includegraphics[width=0.48\textwidth]{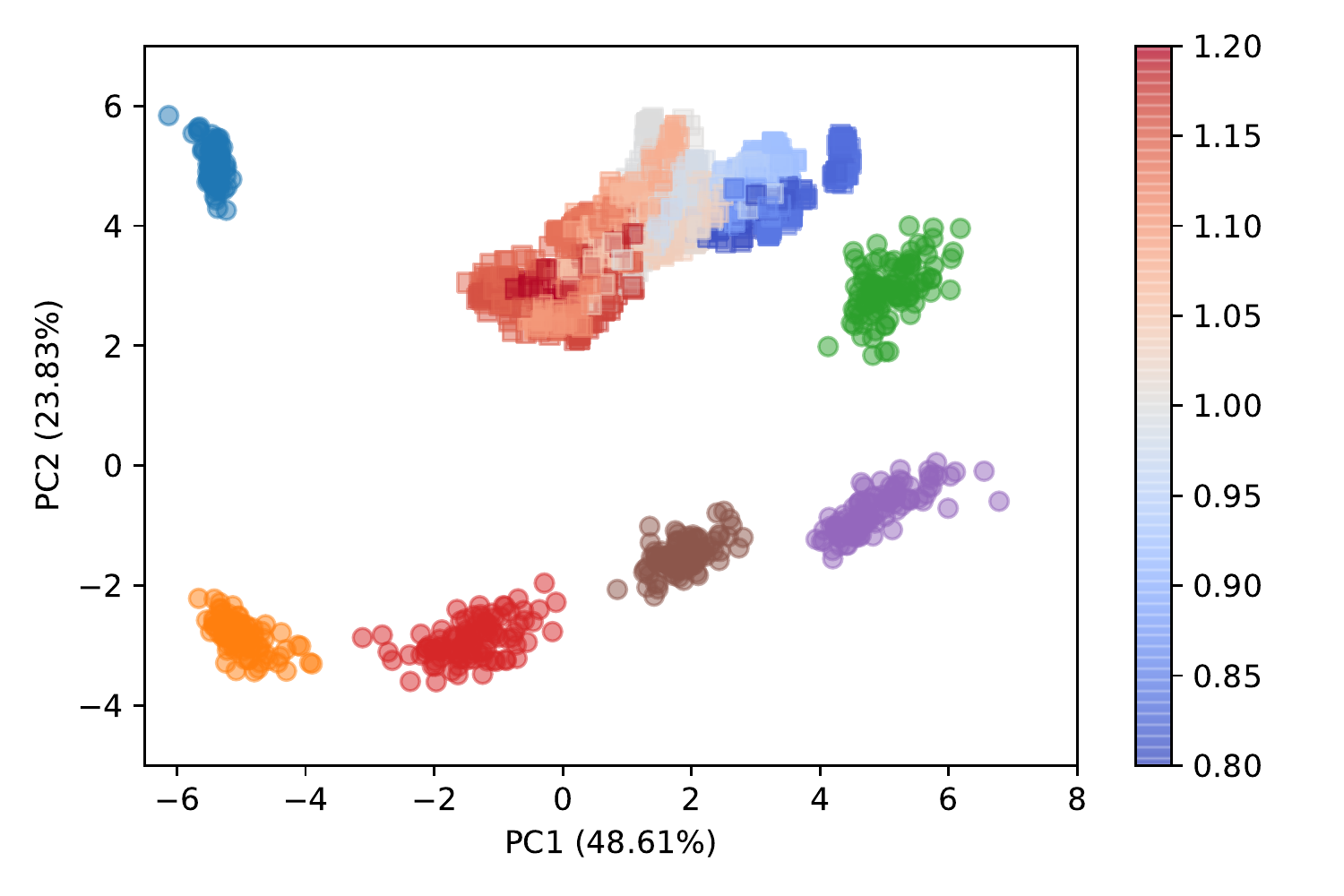}}
    \subfigure[Werckmeister]{\includegraphics[width=0.48\textwidth]{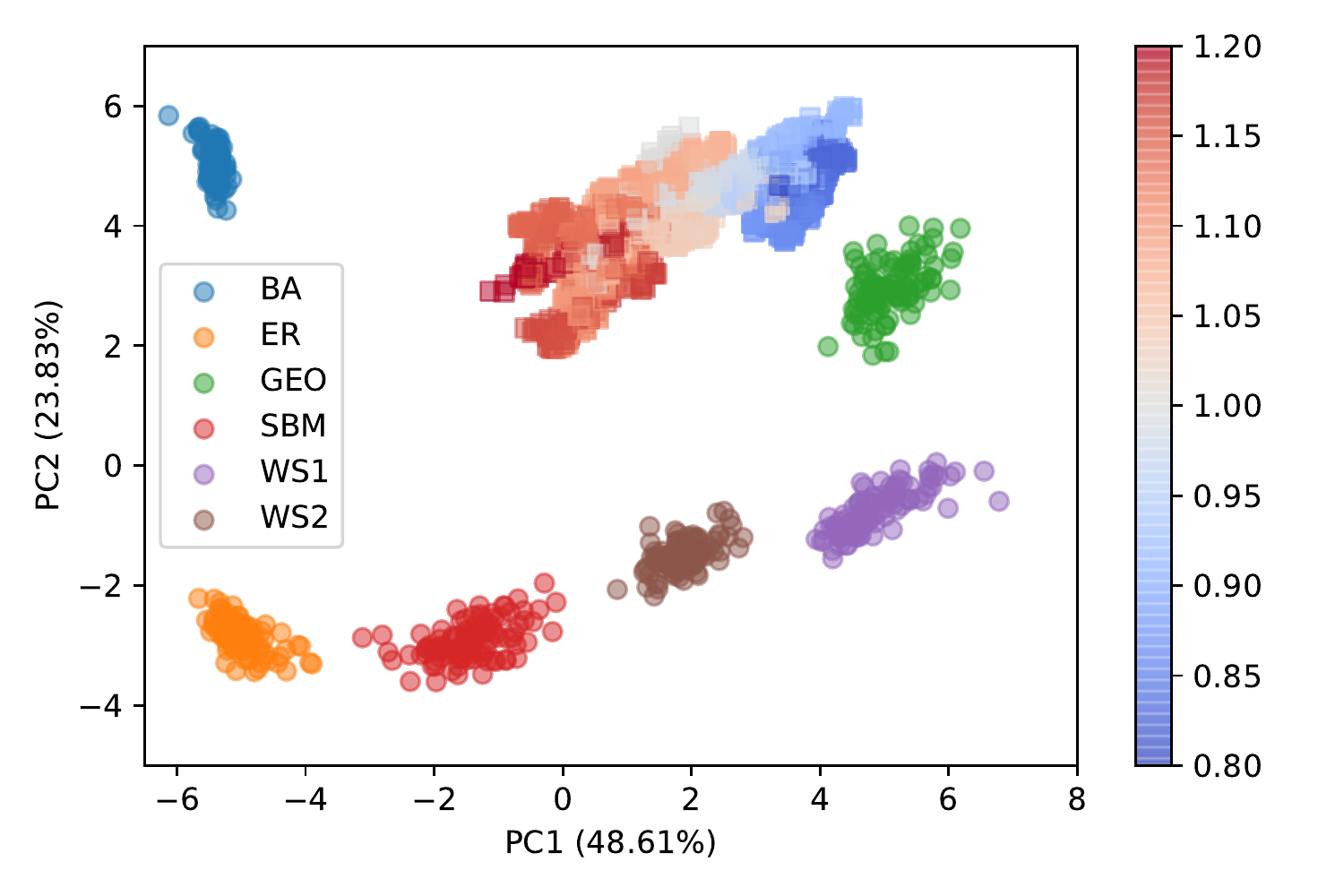}}
    
\caption{PCA projection of the dissonance networks (with shifted partials) for all of the considered temperaments (squares). The circles represent the network models.\label{fig:pcaDissonanceShifted}}
\end{figure*}

The dissonance syntonet obtained for each of the considered temperaments, together with the more traditional complex network models, are shown in Figure~\ref{fig:pcaDissonanceShifted}.

As verified for the consonant syntonets, their dissonance counterparts also tended to populated the space defined between the more traditional models in the PCA space.  The equal temperament networks exhibited the widest dispersion in this space (Figure~\ref{fig:pcaDissonanceShifted}(a)), with the other 4 temperaments yielding more compact syntonet distributions.  Interestingly, the progression of the syntonets in the PCA space exhibited an opposite behavior as observed for the consonance cases. More specifically, we have that the syntonets underwent a relatively smooth shifting in the PCA space in terms of the parameter $\beta$, which was not observed for the equal temperament syntonets.

One of the most interesting results discussed so far regards the impressive ability of the partials-shifted syntonets to produce a wide range of topological organizations, all comprised between the considered more traditional complex network models.  As such, syntonets provide an intrinsic flexibility for generating several topologies, therefore allowing a flexible method for network construction that is completely independent of the musical relationships and implications of the reported syntonets. We also postulate that the suggested methodology to obtain syntonets, especially when considering other types of temperaments not necessarily limited by musical concerns as well anharmonicity transformations other than taking the power of the partials, has the potential to act as a kind of universal generator of networks.  In other words, it is possible that syntonets underlie a general principle of complex network topology.

\begin{figure*}
\centering
\includegraphics[width=0.95\textwidth]{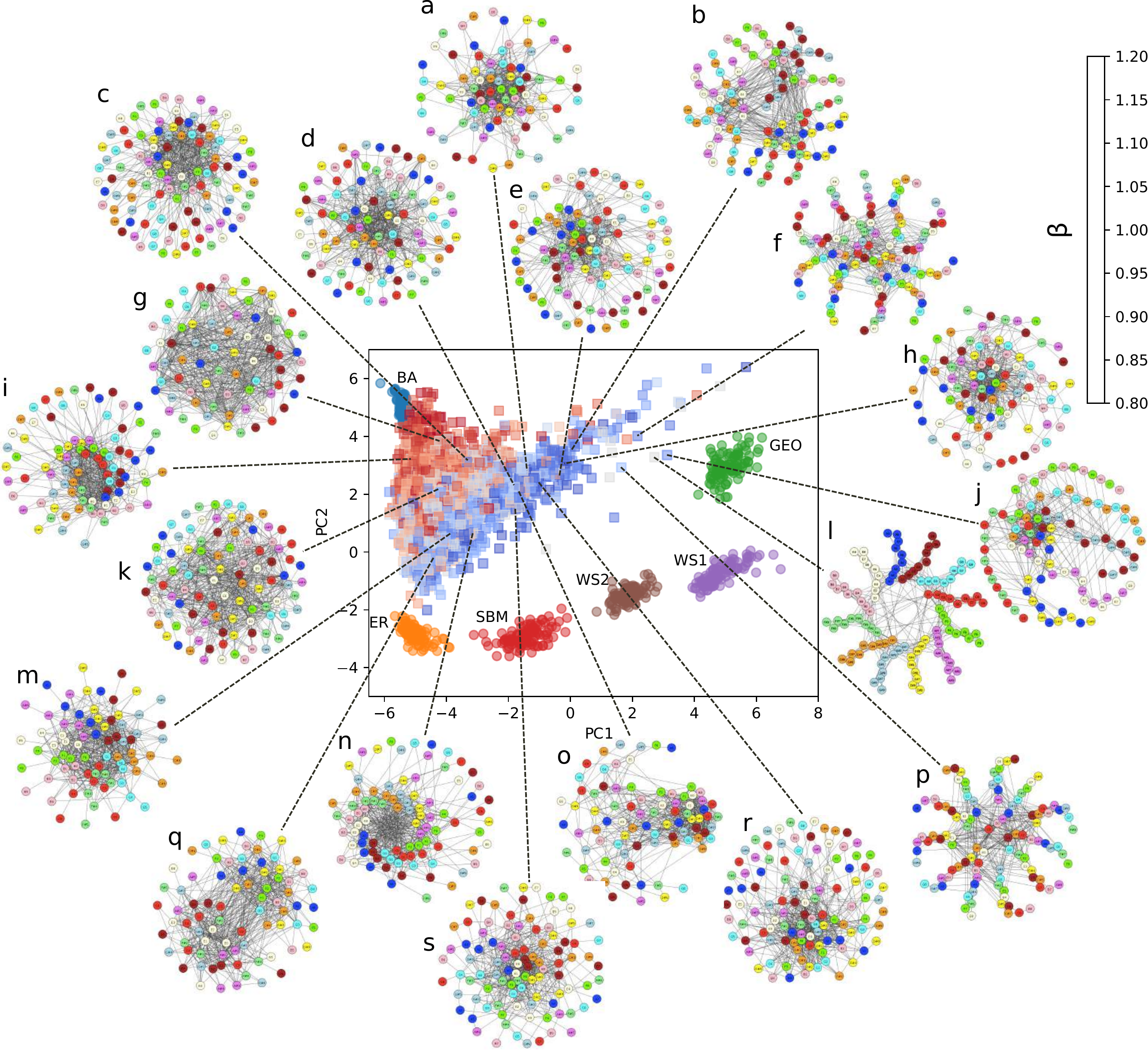}
\caption{Example of syntonet for some selected parameters of shifted partials and the respective positions in PCA. Here, we considered the \emph{equal} temperament. In this projection, PC1 and PC2 represent 48.61\% and 23.83\% of the data variance, respectively.}\label{fig:consonance}
\end{figure*}

In order to better illustrate the topological diversity obtainable from partials-shifted syntonets, we provided the visualization of several networks, which is shown in Figure~\ref{fig:consonance}.  Recall that each network maps into a point in the PCA space.  It is easy to verify from this figure that the syntonets near to the GEO models tended to present a more regular and symmetric topology that, nevertheless, vary substantially even with small increases of $\beta$.  Interestingly, some obtained networks (such as Figure~\ref{fig:consonance}(q)) even presented modular organization (two communities).  Spiral organization can also be noticed in some cases (e.g., Figure~\ref{fig:consonance}(f)~and~(p)).  As the values of $\beta$ increases, the respective networks tend to present hubs and compact cores such as that in Figure~\ref{fig:consonance}(i).

\begin{figure*}
\centering
\includegraphics[width=0.95\textwidth]{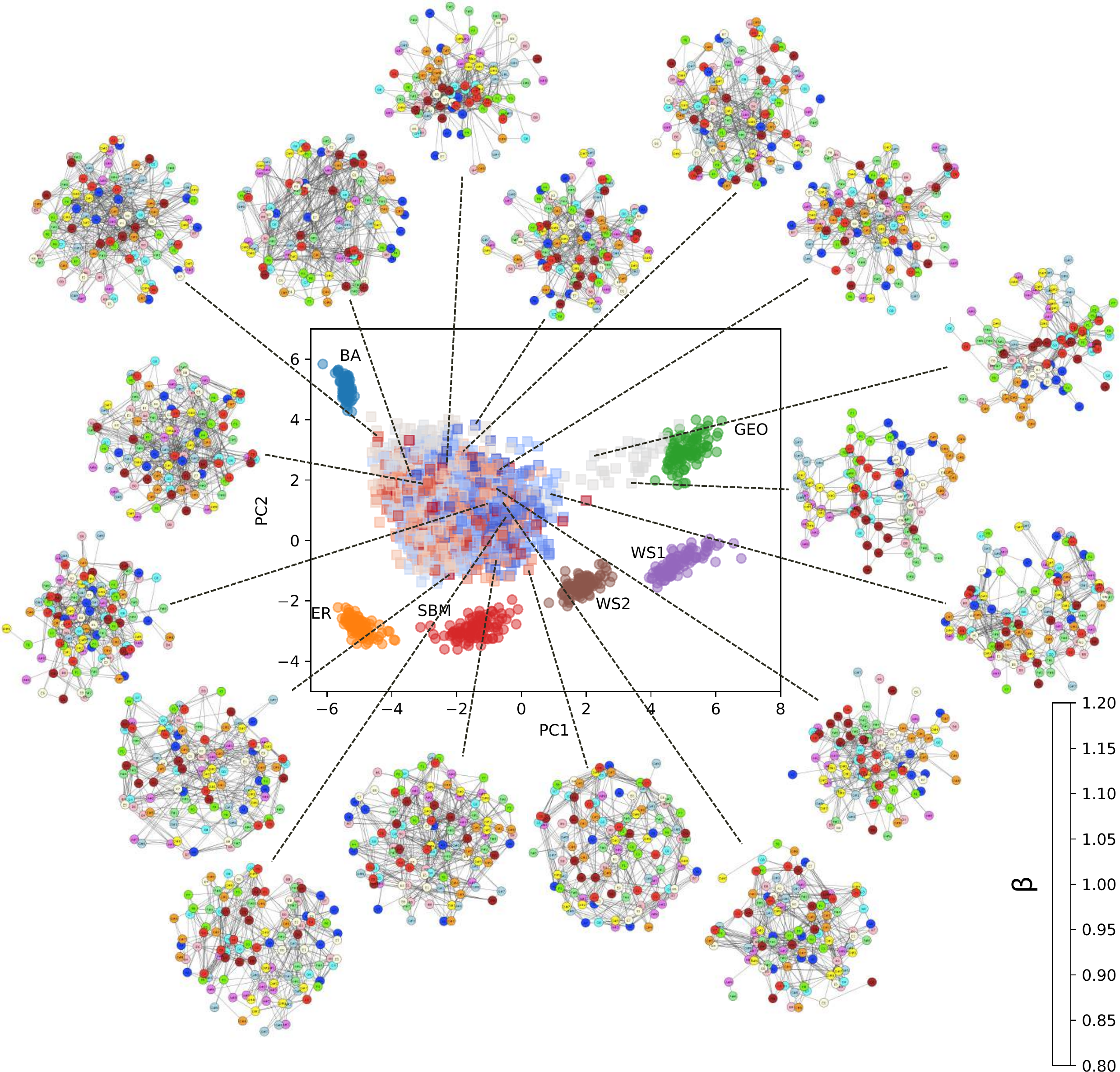}
\caption{Example of consonance networks for some selected parameters of shifted partials and the respective positions in PCA. Here, we considered the \emph{just} temperament. In this projection, PC1 and PC2 represent 48.61\% and 23.83\% of the data variance, respectively.}\label{fig:consonanceJust}
\end{figure*}

Figure~\ref{fig:consonanceJust} shows the visualizations of several consonances partials-shifted syntonets obtained for the just temperament, superimposed on the respective PCA diagram.  The obtained networks have substantial topological diversity, reflected in the respective dispersion in the PCA space. The intrinsic topological differences are more difficult to be appreciated visually.  Similar visualizations were obtained for the Werckmeister, Pythagorean, and Meantone temperaments.

\section{Conclusion}
Complex networks have achieved great importance and popularity, mainly thanks to their ability to represent a wide range of heterogeneous connectivity patterns found in the real world.  In this work, we suggest a procedure to obtain complex networks as induced by consonances and dissonances determined by specific scale modes and temperaments.  The consonances and dissonances were estimated by employing a methodology motivated on Helmholtz's approach, reflecting the alignment between neighboring partials in two given sounds.  Complex networks were obtained by taking into account the obtained levels of consonance and dissonance with respect to five important traditional temperaments while assuming all scales as beginning at $C1$ and extending through five octaves.  The focus was kept on the major mode.  Partial contents considering or not frequency shifting were considered, leading to quite diverse features.

Several results were obtained.  From the musical perspective, we have that the obtained harmonic complex networks (syntonet) that reflected many of the expected characteristics, with the equal temperament yielding particularly regular topological organization.  An interesting result regards the fact that shifted partials can exhibit a remarkable diversity of consonance/dissonance relationships as a consequence of even small variations of $\beta$.  It implies, among other things, that the consonance/dissonance of intervals can vary substantially in musical instruments presenting diverse levels and types of anharmonicity.

Particularly remarkable results have been obtained regarding network science aspects of the obtained syntonet.  It has been shown that shifted partial scales can present an impressive diversity of topological properties that tend to span the region comprised of traditional models such as scale-free, uniform random, geographical, and modular. Syntonets derived from equal temperament resulted particularly interesting as they undergo gradual changes of topology as the anharmonicity parameter $\beta$ is changed. These results suggest that the proposed syntonet could correspond to a universal way to generate networks with diverse topology, especially when considering other temperaments and anharmonicity transformations. It would be related to the almost infinite possibilities of partial alignments implied by the number-theoretical relationships between the obtained partial sequences.  It would be interesting to investigate this possibility further.  Also interesting would be to extend the characterization and comparison of topological properties of the considered networks by taking into account all the obtained topological measurements instead of PCA projections.

\section*{acknowledgement}
Luciano da F. Costa thanks CNPq (grant no. 307085/2018-0) and NAP-PRP-USP for sponsorship. Henrique F. de Arruda acknowledges FAPESP for sponsorship (grant no. 2018/10489-0 and no. 2019/16223-5). This work has been supported also by FAPESP grants 2015/22308-2.

\bibliographystyle{ieeetr}
\bibliography{references}

\end{document}